\documentclass[conference]{IEEEtran}
\IEEEoverridecommandlockouts
\usepackage{cite}
\usepackage{amsmath,amssymb,amsfonts}
\usepackage{algorithmic}
\usepackage{graphicx}
\usepackage{textcomp}
\usepackage{xcolor}
\usepackage{multirow}
\usepackage{svg}
\usepackage{enumitem}
\usepackage{subcaption}
\usepackage{booktabs}
\usepackage[numbers,sort&compress]{natbib}
\def\BibTeX{{\rm B\kern-.05em{\sc i\kern-.025em b}\kern-.08em
    T\kern-.1667em\lower.7ex\hbox{E}\kern-.125emX}}
\begin{document}

\title{Multi-Channel Masking with Learnable Filterbank for Sound Source Separation
}
\author{\IEEEauthorblockN{Wang Dai, Archontis Politis, Tuomas Virtanen}
\IEEEauthorblockA{\textit{Audio Research Group} \\
\textit{Tampere University}\\
Tampere, Finland \\
\{wang.dai, archontis.politis, tuomas.virtanen\}@tuni.fi}}

\maketitle

\begin{abstract}
This work proposes a learnable filterbank based on a multi-channel masking framework for multi-channel source separation. The learnable filterbank is a 1D Conv layer, which transforms the raw waveform into a 2D representation. In contrast to the conventional single-channel masking method, we estimate a mask for each individual microphone channel. The estimated masks are then applied to the transformed waveform representation like in the traditional filter-and-sum beamforming operation. Specifically, each mask is used to multiply the corresponding channel’s 2D representation, and the masked output of all channels are then summed. At last, a 1D transposed Conv layer is used to convert the summed masked signal into the waveform domain.  The experimental results show our method outperforms single-channel masking with a learnable filterbank and can outperform multi-channel complex masking with STFT complex spectrum in the STGCSEN model if a learnable filterbank is transformed to a higher feature dimension. The spatial response analysis also verifies that multi-channel masking in the learnable filterbank domain has spatial selectivity.
\end{abstract}


\section{Introduction}
Sound source separation has many applications, which include for example signal enhancement for automatic speech recognition and hearing-aids. In recent years, deep learning-based single-channel source separation has been largely investigated with impressive results, e.g., \cite{luo2018tasnet, Isik2016SingleChannelMS, chen2017deep, takahashi2017multi, liu2020deep}. These methods are often conducted in the time-frequency domain. However, accurate reconstruction of the phase of the clean sources is a nontrivial problem and the erroneous estimation of the phase introduces an upper bound on the accuracy of the reconstructed audio \cite{luo2018tasnet}. Models that use learned filterbank such as TasNets are able to consistently outperform models based on the short-time Fourier transform (STFT) features in speech separation \cite{luo2018tasnet, luo2019conv, bahmaninezhad2019comprehensive}. 

In the multi-channel scenario, separation performance can be improved by leveraging the additional spatial information between different microphone signals. Therefore, the extraction of spatial features is key to the success of multi-channel source separation. Conventionally, multi-channel signal enhancement or separation is often handled through beamforming (or spatial filtering) \cite{benesty2008microphone, Markovich-Golan2018b}, which applies a linear spatial filter to weight (mask) different microphone channels in the STFT domain in order to suppress non-target source signal components. However, its performance depends on reliable estimation of spatial information, which can be challenging in noisy conditions.

Motivated by the success of deep learning in single-channel source separation, a recent line of works propose to combine supervised single-channel techniques with beamforming methods for multi-channel cases, e.g., \cite{erdogan16_interspeech, 8691791, wang2018all,luo2019fasnet}. These approaches are often termed neural beamforming, which is mainly accomplished by predicting the beamforming weights directly or indirectly. More recently, all-deep-learning beamforming paradigms, e.g., ADL-MVDR \cite{Zhang2020ADLMVDRAD}, EaBNet \cite{li2022embedding}, TaylorBeamformer \cite{li22c_interspeech}, and \cite{9747528} were proposed. In these works, the mask estimation, spatial covariance calculation, and beamforming are integrated into a whole network and trained in an end-to-end manner. The authors in \cite{ ren2021causal} proposed to estimate multi-channel masks where each mask weights (masks) the noisy signal of the corresponding input channel, then the masked signals are summed, acting as a filter-and-sum beamforming operation. This method is simple and effective while it does not require additional spatial features such as the inter-channel phase and level differences. However, this method operates in the STFT domain, and similar operations with a learnable filterbank have not been exploited.

In this paper, we explore the multi-channel masking strategy in the transformed waveform representation (i.e., learnable filterbank domain) for multi-channel sound source separation. Specifically, we use one-dimensional convolution (1D Conv) layer to encode each channel's time-domain signal into a 2D intermediate representation as the input of the multi-mask estimation network. After estimating multiple masks, the similar process is performed to that of the traditional filter-and-sum beamforming, i.e., each channel mask is element-wise multiplicated with the output of the filterbank, and the masked (weighted) representations are then summed. 
We evaluate the approach using a simulated dataset on two source separation models based on IC Conv-TasNet \cite{lee2021inter} and STGCSEN \cite{9746054}. 
\begin{figure}[htbp]
  \centering
  \includegraphics[width=1\linewidth]{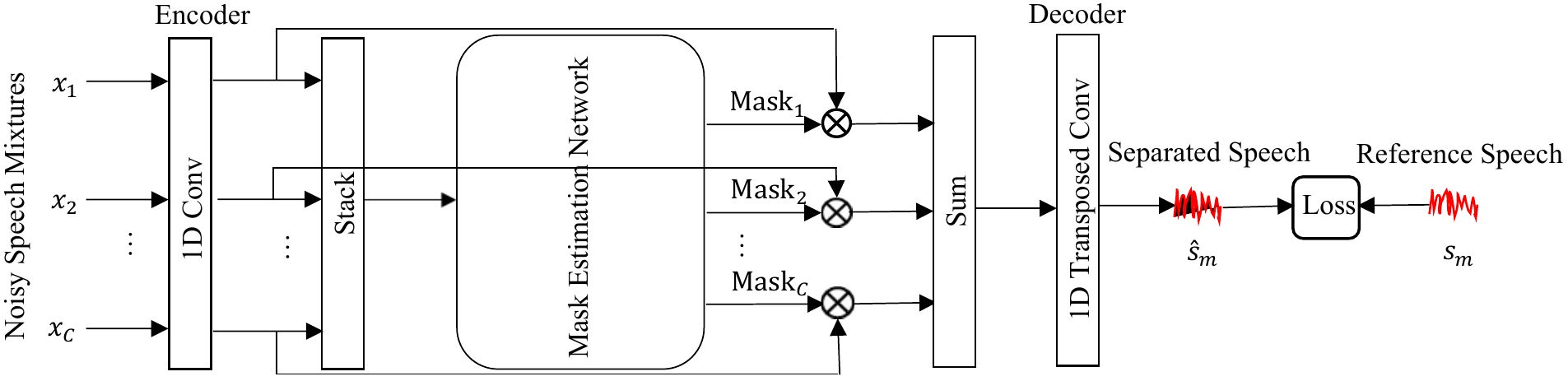}
  \caption{Overview of proposed multi-channel masking framework for sound source separation}
  \label{fig:overviewOfFramework}
\end{figure}
\section{Multi-channel masking with learnable filterbank}
We consider multi-channel speech and noise mixture signal of $c=1,..,C$ channels in the time domain:
\begin{equation}
{x}_{c}(t) = {s}_{c}(t) + {n}_{c}(t),
\end{equation}
where ${x}_{c}(t)$, ${s}_{c}(t)$, and ${n}_{c}(t)$ are the noisy speech, clean speech and noise captured by the $c$-th microphone at time $t$, respectively. 
Our aim is to extract the speech signal ${s}_{m}(t)$, while suppressing the noise component ${n}_{m}(t)$, i.e., to recover one single-channel clean speech signal, where ${m}$ denotes reference microphone channel.
The proposed architecture for multi-channel sound source separation is shown in Fig. 1. It consists of three modules: learnable encoder, multi-channel masking module and learnable decoder, all of which are described in subsequent sections. 
\subsection{Learnable encoder}
Similarly to IC Conv-TasNet, we segment the signal into segments which length is $T$ samples ${x}_{c}$, denoted as ${x}_{c} \in {\mathbb R} ^ {T \times N}$. $N$ represents the number of segments, i.e., time frames. We then use a 1D Conv layer with $F$ filters to encode these segmented waveform signals into 2D intermediate representation, but without ReLU activation function. The transformation is formulated as
\begin{equation}
{X}_{c} = {U}_{c} \cdot {x}_{c},
\end{equation}
where ${U}_{c} \in {\mathbb R} ^ {F \times T}$ is the 1D Conv filters with ${F \times T}$ learnable parameters and $\cdot$ represents the multiplication operation. ${X}_{c} \in {\mathbb R} ^ {F \times N}$ is the transformed 2D representation. The 1D Conv encoder converts each segmented waveform signal into a feature vector of length $F$. 

\subsection{Multi-channel masking module}
Conventional single-channel masking networks estimate single mask (real-valued or complex-valued) for multi-channel input signals, which is used to multiply reference microphone to estimate the desired signal. However, the mask is only estimated for the reference channel. To utilize full-channel information, \cite{9747528, ren2021causal} proposed learning-based deep beamforming networks, which use neural networks to predict the complex-valued mask for each channel. With multi-channel inputs, the beamforming network filters the multi-channel STFT of array signals to produce enhanced signals.

Motivated by the success of the frequency domain neural filter-and-sum beamforming method in speech enhancement, we borrow the neural filter-and-sum beamforming idea to learnable filterbank domain for source separation.
As seen in Fig. 1, the mask estimation network estimates the mask for each encoder channel, and the mask values can be freely estimated. 
We will summarize possible multi-mask estimation networks in Section III. 
After getting multiple masks, each estimated mask is element-wise multiplicated with the corresponding channel’s encoder output. The final filterbank domain representation of the reference channel $\hat{S}_{m} \in {\mathbb R} ^ {F \times N}$ is obtained as 
\begin{equation}
\hat{S}_{m} = \sum_{c=1}^C {M}_{c} \odot {X}_{c}, 
\label{eq:masking}
\end{equation}
where ${M}_{c} \in {\mathbb R} ^ {F \times N}$ indicates the mask of the $c$-th channel and $\odot$ represents element-wise multiplication. This process can be viewed as filter-and-sum beamforming operation and ${M}=\{{M}_c\}_{c=1}^C \in {\mathbb R} ^ {C \times F \times N}$ can be taken as the weights of a filter-and-sum beamformer in the transform domain.

\subsection{Learnable decoder}
After getting the masked encoder outputs $\hat{S}_{m}$, a 1D transposed Conv is used to reconstruct them into time domain signal. The segments of extracted target speech are obtained as 
\begin{equation}
\hat{s}_{m} = {V} \cdot \hat{S}_{m},
\label{eq:decoding}
\end{equation}
where ${V} \in {\mathbb R} ^ {T \times F}$ is the 1D transposed Conv filters with ${T \times F}$ learnable parameters. The reconstructed segments $\hat{s}_{m} \in {\mathbb R} ^ {T\times N}$ are transformed into the final waveform using the overlap-and-add operation. 

\section{Evaluation}

\subsection{Dataset}
The dataset is constructed based on the publicly available CHiME-3 simulated dataset \cite{barker2015third}.
Since there is a mismatch between training, development and testing sets \cite{liu2020multichannel, li2019multichannel}, we only use the CHiME-3 simulated training set and generate new training, validation and testing subsets. The CHiME-3 simulated training data was constructed by mixing clean utterances into environment background recordings.

The multi-channel background noises were recorded in four noisy locations: bus, cafe, pedestrian area, and street junction. For each location, four or five sessions were recorded at different time, with a duration of about 0.5 hours for each session. Every noise location has different sessions and the noises distribute uniformly in a single session. We cut the long noise sessions into many short non-overlapping samples for each channel according to the maximal speech sample duration of the WSJ0 SI-84 training set. Those short noise samples of every long noise session were randomly divided to three noise partitions: training, validation and test. The training, validation and test noise partitions have approximately 60\%, 20\% and 20\% samples, respectively. We also split the original clean speech samples into training, validation and test subsets according to different speakers, resulting in 4500, 1388 and 1250 speech samples, respectively. The clean speech samples were then convolved with time-varying impulse responses and the generated multi-channel speech samples were used as the target speech signals to mixture with noise samples. 
The time-varying impuse response is computed in the STFT domain using a multiplicative transfer function between microphone array recording and a simultaneous close-miked recording of CHiME-3 real single-speaker recordings. 
The signal powers of the multi-channel speech samples were then normalized to match target SNR values, with respect to the total power of the multi-channel noise recording across channels. The SNRs between speech and noise signals have an average of approximately 5 dB. As the short noise samples have longer duration than most of speech samples, we randomly choose each noise sample's starting and ending point while keeping it the same duration as the target speech sample for every channel. 
The chosen noise sample was then added to the normalized speech sample for each channel to get multi-channel noisy speech mixtures. Some of the noise samples might be selected multiple times for the number of noise samples is less than speech samples in every partition. The new training, validation and test sets have 4500, 1388 and 1250 noisy speech samples in each channel, respectively.

\subsection{Baseline methods and evaluation models}
We compare our method to two baselines: single-channel masking with learnable filterbank and multi-channel complex masking with STFT complex spectrum. We use the same multi-channel complex masking with STFT complex spectrum approach as described in \cite{ren2021causal}. The performances of these methods are examined in two different model architectures for multi-channel source separation based on IC Conv-TasNet and STGCSEN. 
In the original IC Conv-TasNet, following the last PReLU activation function, two ${1\times1}$ 2D Conv layers are used to map channel dimension to one and feature dimension to encoder dimension, respectively. A sigmoid function is subsequently used to predict a mask.
In the proposed method, we use $C$ ${1\times1}$ 2D Conv layers in parallel and every 2D Conv layer maps the channel dimension to one, thus we get six independent output channels in total. 
Each of the parallel six 2D Conv layers is followed by one 1 x 1 2D Conv layer to map the number of features to each channel’s encoder dimension. Since the multi-channel masking method performs like a beamformer, beamformer weights should not be limited to [0-1]. We directly take the output of each channel as the mask instead of using the sigmoid function. 
When estimating a single-channel mask based on IC Conv-TasNet, we remove the sigmoid function to output a real-valued mask to make it comparable with multi-channel masking. 

For STGCSEN-based model, we take five spatial-temporal modules of STGCSEN as the mask estimation network to estimate single mask or multiple masks. More specifically, we use a ${1 \times 1}$ 2D Conv layer to reduce channel dimension to one from the last spatial-temporal module, then estimate a real-valued mask. In the case of estimating multiple masks, we use six ${1 \times 1}$ 2D Conv layers to map channel dimension to six from the last spatial-temporal module.
When comparing multi-channel masking with different representations, we only need to switch between 1D Conv encoder/decoder and STFT/inverse STFT.  

\subsection{Training and network configuration}
The loss function to minimize was negative signal-to-distortion ratio (SDR) \cite{vincent2006performance}. We use the Adam optimizer to train the models with a fixed learning rate of ${3 \times 10 ^ {- 3}}$ and total of 50 training epochs with a mini-batch of size eight utterances. The model with the most smallest loss on the validation set was selected as the final model. In training, we cut the data into 3-second segments. If the length of an utterance is longer than 3 seconds, it is cut to 3 seconds, and if it is shorter than 3 seconds, it is made to 3 seconds by zero padding.
We selected hyperparameters of the best performing IC Conv-TasNet \cite{lee2021inter}: 8 dilated Conv layers in each TCN blocks, 3 TCN blocks, 512 encoder dimension, 128 bottleneck feature dimension, 64 input channel dimension in dilated Conv layers, and 256 output channel dimension in dilated Conv layers.


\subsection{Results}
\begin{table}[thbp]
  \centering  
  \caption{Performance of IC Conv-TasNet}
  \label{tab:IC ConvTasNet}
  \centering
  \begin{tabular}{l c c c c c c}
    \toprule
    \multicolumn{1}{l}{\textbf{Methods}} & \multicolumn{1}{c}{\textbf{$F$}} & \multicolumn{1}{c}{\textbf{$T$}} & \multicolumn{1}{c}{\textbf{PESQ}} & \multicolumn{1}{c}{\textbf{STOI}} & \multicolumn{1}{c}{\textbf{SDR}}\\
    \midrule
    \multirow{1}{*}{SC-Learn} 
     &512 &256 &2.66 &0.963 &19.7 \\ \hline
    \multirow{3}{*}{MC-Learn} 
    &512 &256 &2.68 &0.964 &19.9 \\
    &256 &256 &2.72 &0.966 &20.0 \\
    &512 &512 &2.40 &0.954 &18.4 \\ \hline
    \multirow{2}{*}{MC-STFT} 
    &256 &256 &2.86 &0.969 &20.7 \\
    &512 &512 &2.62 &0.963 &19.6 \\ 
    \bottomrule
  \end{tabular} 
\end{table}

\begin{table}[thbp]
  \centering  
  \caption{Performance of STGCSEN}
  \label{tab:STGCSEN}
  \centering
  \begin{tabular}{l c c c c c}
    \toprule
    \multicolumn{1}{l}{\textbf{Methods}} & \multicolumn{1}{c}{\textbf{$F$}} & \multicolumn{1}{c}{\textbf{$T$}} & \multicolumn{1}{c}{\textbf{PESQ}} & \multicolumn{1}{c}{\textbf{STOI}} & \multicolumn{1}{c}{\textbf{SDR}}\\
    \midrule
    \multirow{1}{*}{SC-Learn} 
    &512 &256 &2.23 &0.938 &16.0 \\ \hline
    \multirow{3}{*}{MC-Learn} 
    &512 &256 &2.34 &0.952 &17.8 \\
    &256 &256 &2.05 &0.947 &16.6 \\
    &512 &512 &2.10 &0.944 &16.6 \\ \hline
    \multirow{2}{*}{MC-STFT} 
    &256 &256 &2.06 &0.946 &16.4 \\
    &512 &512 &2.25 &0.950 &17.0 \\ 
    \bottomrule
  \end{tabular} 
\end{table}
Three metrics were used to measure the separation performance: SDR, perceptual evaluation of speech quality (PESQ) \cite{rix2001perceptual}, and the short-time objective intelligibility measure (STOI) \cite{taal2011algorithm}.
Table 1 and Table 2 present the performance of IC Conv-TasNet and STGCSEN based separation models in test set under different $F$ and $T$ values. $F$ represents the number of encoder features or frequency bins of STFT complex spectrum (real+imaginary). $T$ indicates the window size. The hop size is set to half of the window size.
We make a comparison between multi-channel masking with learnable filterbank (indicated as MC-Learn) and single-channel masking with learnable filterbank (indicated as SC-Learn) with ($F$, $T$) = (512, 256) for both models. MC-Learn method outperforms the SC-Learn method in all three objective metrics, especially for the STGCSEN model. It demonstrates that multi-channel masking can learn more spatial information, acting as a beamformer.
For multi-channel masking with learnable filterbank vs. multi-channel complex masking with STFT complex spectrum (indicated as MC-STFT), we set two groups of ($F$, $T$): (256, 256) and (512, 512). From the testing results, We can see MC-STFT method generally has better performance than the MC-Learn method when $F$ equals $T$. That might be induced by the benefit of complex masking, which can make more use of phase information. Except for evaluating the two methods in the same $F$ and $T$ values, we also compare different different $F$ and $T$ values, i.e., (512, 256). We find an interesting result, in STGCSEN model, MC-Learn method outperforms the MC-STFT method when $F$ is larger than $T$. That could be attributed to a larger feature dimension has more discriminative representation for separation when the learnable filterbank transforms the windowed waveform signal into large feature space. 

\section{Analysis of spatial response}
The multi-channel masking solution with learnable filterbank can be seen as a signal-dependent beamforming operation working in a transform domain other than the typical Fourier one. That allows us to analyze also the spatial properties of the masks. 
However, multi-channel masks in a time-frequency domain such as the STFT have a direct interpretation as beamforming weights for a certain time-frequency point, and spatial analysis can be performed by applying such weights to the modeled or measured multi-channel array frequency response for multiple directions. In the signal domain of a learnable filterbank, such a direct interpretation is not possible. To analyze the spatial properties of the masking network in this case, we treat it as a black box and we observe its input-output signal relationships by passing appropriate time-domain input signals.
Our evaluation uses the generated simulated test set, and our process to evaluate the spatial responses consists of the following steps:
\begin{enumerate}[label=\alph*.]
\item We model the steering vectors $\mathbf{h}_k(n) = [h_1^{(k)}(n), ..., h_C^{(k)}(n)]$ of the 6-channel array under study for a dense uniform spherical grid of $K=5100$ directions-of-arrival (DOA), using the array geometry of the CHiME-3. We assume omnidirectional sensors and plane-wave propagation, in which case the steering vectors reduce to pure delay terms dependent on each DOA. 
\item We simulate discrete sinusoidal input signals $z(f,n) = \sin 2\pi f n/F_\mathrm{sr}$ of frequency $f$ for the target samplerate $F_\mathrm{sr} = 16$ kHz with the same duration as the speech utterances in the dataset. These are convolved with the array steering vectors for each $k$th direction, resulting in multi-channel sinusoidal array signals $\mathbf{x}_k(f,n) = z(f,n)*\mathbf{h}_k(n)$, $k=1,..,K$, with $*$ denoting convolution operation.
\item To simplify the analysis further, the time-varying position of the CHiME-3 single speaker recordings is extracted using the provided speaker localization routines of CHiME-3, and recordings are selected in which the speaker position does not vary significantly over the duration of the utterance (in the range of 2-3 cm). These recordings are used to generate reference multi-channel masks that are used for the spatial analysis, with a known mean speaker position and DOA.
\item The sinusoidal recordings $\mathbf{x}_k(f,n)$ are passed through the encoder-decoder network, and a reference multi-channel mask is applied to them, computed in the previous step. The output signal of the network $y(f,n)$ is then generated through Eq.~(\ref{eq:masking})-(\ref{eq:decoding}).
\item By analyzing the input-output power ratio we evaluate a spatial beamforming power response $b_k(f) = \sum_n y^2(f,n)/\sum_n z^2(f,n)$ , for excitation frequency $f$. 
\end{enumerate}

For a certain reference speaker DOA from one of the test samples resulting in reference masks applied to the input signals, we expect to see a spatial response that is concentrated around the speaker DOA. In Fig. 2, we can see one such example for a speaker at a position around 25 cm above/front of the array, resulting in a DOA of (156°, 86°) in azimuth and elevation. The red dot indicates the speaker DOA. Fig. 2 (a) shows the beampattern of a plain delay-and-sum beamformer steered towards the known speaker DOA, while Fig. 2 (b) shows the beampattern of multi-channel masking with a learnable filterbank (($F$, $T$)=(512, 256)). In this case the DOA is unknown, however the masking clearly indicates spatial selectivity around the speaker DOA.
´

\begin{figure}[t]
\centering
\begin{subfigure}{.5\linewidth}
  \centering
  \includegraphics[width=.99\linewidth]{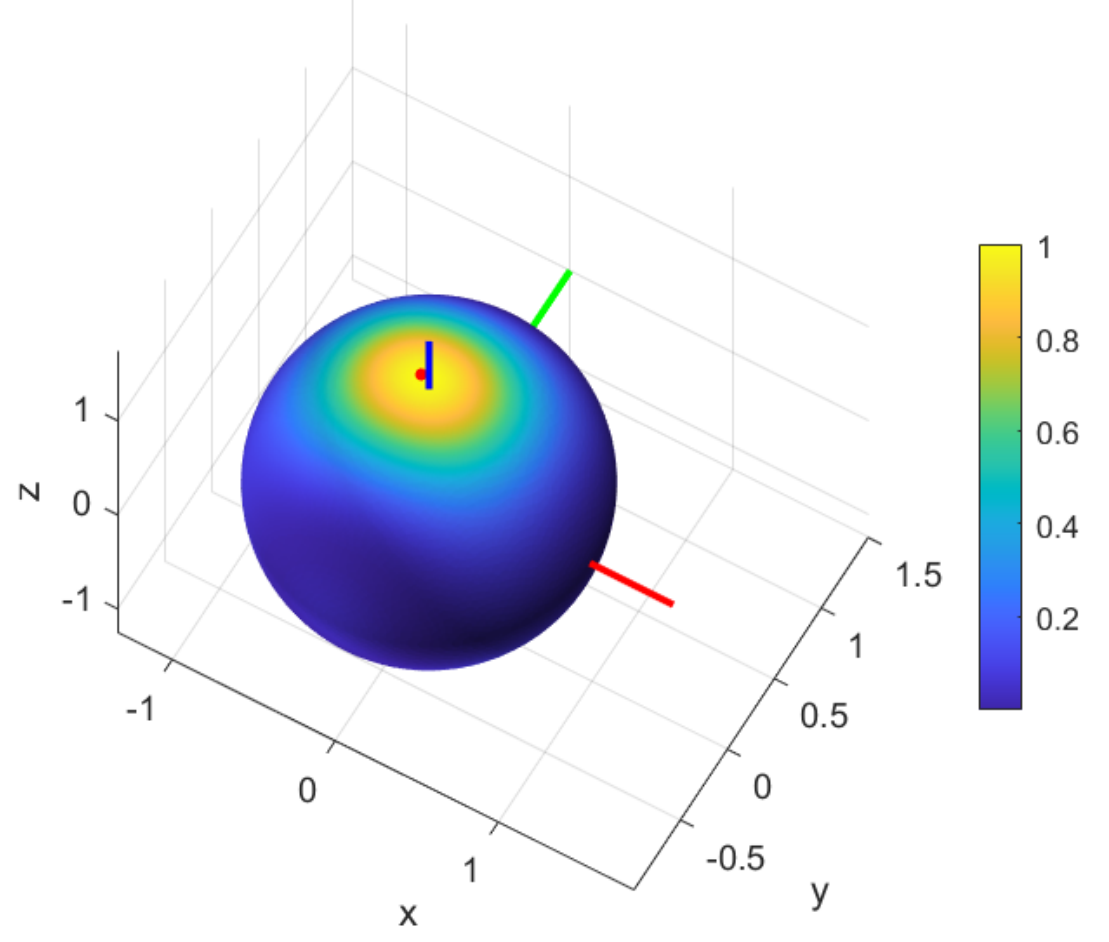}
  \caption{}
  \label{fig:sub1}
\end{subfigure}%
\begin{subfigure}{.5\linewidth}
  \centering
  \includegraphics[width=.99\linewidth]{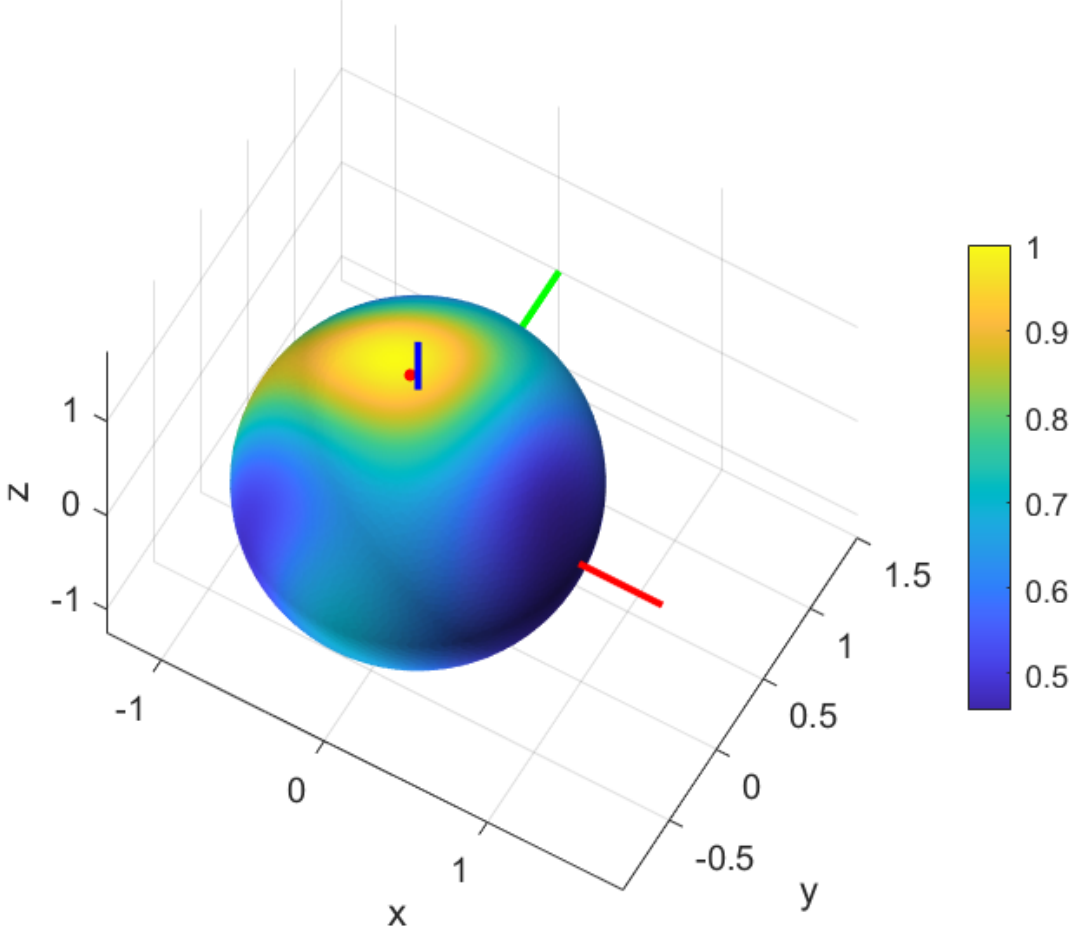}
  \caption{}
  \label{fig:sub2}
\end{subfigure}
\caption{Beampattern for sinusoidal input signal of 1kHz: (a) Delay-and-sum beamformer in the time domain steered to speaker DOA, (b) Multi-channel masking with learnable filterbank domain. The red dot indicates the target speaker DOA.}
\label{fig:test}
\end{figure}
\section{Conclusions}
In this study, we proposed a novel neural spatial filtering method based on multi-channel masking with a learnable filterbank, which essentially estimates a group of ﬁlters and applies ﬁlter-and-sum beamforming operation. We evaluated it by separating the reference channel's speech signal from multi-channel noisy speech mixtures. The experimental results show the proposed method can outperform single-channel masking with learnable filterbank approach and also performs better than multi-channel complex masking with STFT complex spectrum in the STGCSEN model. The analyse of spatial response demonstrates our approach has indeed spatial selectivity. 

\bibliographystyle{IEEEtran}
\bibliography{conference}

\begin{thebibliography}{10}
\providecommand{\url}[1]{#1}
\csname url@samestyle\endcsname
\providecommand{\newblock}{\relax}
\providecommand{\bibinfo}[2]{#2}
\providecommand{\BIBentrySTDinterwordspacing}{\spaceskip=0pt\relax}
\providecommand{\BIBentryALTinterwordstretchfactor}{4}
\providecommand{\BIBentryALTinterwordspacing}{\spaceskip=\fontdimen2\font plus
\BIBentryALTinterwordstretchfactor\fontdimen3\font minus
  \fontdimen4\font\relax}
\providecommand{\BIBforeignlanguage}[2]{{%
\expandafter\ifx\csname l@#1\endcsname\relax
\typeout{** WARNING: IEEEtran.bst: No hyphenation pattern has been}%
\typeout{** loaded for the language `#1'. Using the pattern for}%
\typeout{** the default language instead.}%
\else
\language=\csname l@#1\endcsname
\fi
#2}}
\providecommand{\BIBdecl}{\relax}
\BIBdecl

\bibitem{luo2018tasnet}
Y.~Luo and N.~Mesgarani, ``{TasNet}: time-domain audio separation network for
  real-time, single-channel speech separation,'' in \emph{2018 IEEE
  International Conference on Acoustics, Speech and Signal Processing
  (ICASSP)}, 2018, pp. 696--700.

\bibitem{Isik2016SingleChannelMS}
Y.~Z. Isik, J.~L. Roux, Z.~Chen, S.~Watanabe, and J.~R. Hershey,
  ``Single-channel multi-speaker separation using deep clustering,'' in
  \emph{Interspeech}, 2016.

\bibitem{chen2017deep}
Z.~Chen, Y.~Luo, and N.~Mesgarani, ``Deep attractor network for
  single-microphone speaker separation,'' in \emph{2017 IEEE International
  Conference on Acoustics, Speech and Signal Processing (ICASSP)}, 2017, pp.
  246--250.

\bibitem{takahashi2017multi}
N.~Takahashi and Y.~Mitsufuji, ``Multi-scale multi-band densenets for audio
  source separation,'' in \emph{2017 IEEE Workshop on Applications of Signal
  Processing to Audio and Acoustics (WASPAA)}, 2017, pp. 21--25.

\bibitem{liu2020deep}
Y.~Liu, M.~Delfarah, and D.~Wang, ``Deep casa for talker-independent monaural
  speech separation,'' in \emph{ICASSP 2020-2020 IEEE International Conference
  on Acoustics, Speech and Signal Processing (ICASSP)}, 2020, pp. 6354--6358.

\bibitem{luo2019conv}
Y.~Luo and N.~Mesgarani, ``{Conv-TasNet}: Surpassing ideal time--frequency
  magnitude masking for speech separation,'' \emph{IEEE/ACM Transactions on
  Audio, Speech, and Language Processing}, vol.~27, no.~8, pp. 1256--1266,
  2019.

\bibitem{bahmaninezhad2019comprehensive}
F.~Bahmaninezhad, J.~Wu, R.~Gu, S.-X. Zhang, Y.~Xu, M.~Yu, and D.~Yu, ``A
  comprehensive study of speech separation: Spectrogram vs waveform
  separation,'' \emph{Proc. Interspeech 2019}, pp. 4574--4578, 2019.

\bibitem{benesty2008microphone}
J.~Benesty, J.~Chen, and Y.~Huang, \emph{{Microphone Array Signal
  Processing}}.\hskip 1em plus 0.5em minus 0.4em\relax Springer Science \&
  Business Media, 2008, vol.~1.

\bibitem{Markovich-Golan2018b}
S.~Markovich-Golan, W.~Kellermann, and S.~Gannot, \emph{Audio Source Separation
  and Speech Enhancement}.\hskip 1em plus 0.5em minus 0.4em\relax Wiley, Sep.
  2018, ch. Spatial filtering.

\bibitem{erdogan16_interspeech}
H.~Erdogan, J.~R. Hershey, S.~Watanabe, M.~I. Mandel, and J.~L. Roux,
  ``{Improved MVDR Beamforming Using Single-Channel Mask Prediction
  Networks},'' in \emph{Proc. Interspeech 2016}, 2016, pp. 1981--1985.

\bibitem{8691791}
S.~Chakrabarty and E.~A.~P. Habets, ``Time–frequency masking based online
  multi-channel speech enhancement with convolutional recurrent neural
  networks,'' \emph{IEEE Journal of Selected Topics in Signal Processing},
  vol.~13, no.~4, pp. 787--799, 2019.

\bibitem{wang2018all}
Z.-Q. Wang and D.~Wang, ``All-neural multi-channel speech enhancement.'' in
  \emph{Interspeech}, 2018, pp. 3234--3238.

\bibitem{luo2019fasnet}
Y.~Luo, C.~Han, N.~Mesgarani, E.~Ceolini, and S.-C. Liu, ``{FaSNet}:
  {Low-latency adaptive beamforming for multi-microphone audio processing},''
  in \emph{{2019 IEEE Automatic Speech Recognition and Understanding Workshop}
  (ASRU)}, 2019, pp. 260--267.

\bibitem{Zhang2020ADLMVDRAD}
Z.~Zhang, Y.~Xu, M.~Yu, S.-X. Zhang, L.~Chen, and D.~Yu, ``{ADL-MVDR: All Deep
  Learning MVDR Beamformer for Target Speech Separation},'' \emph{ICASSP 2021 -
  2021 IEEE International Conference on Acoustics, Speech and Signal Processing
  (ICASSP)}, pp. 6089--6093, 2020.

\bibitem{li2022embedding}
A.~Li, W.~Liu, C.~Zheng, and X.~Li, ``Embedding and beamforming: All-neural
  causal beamformer for multichannel speech enhancement,'' in \emph{ICASSP
  2022-2022 IEEE International Conference on Acoustics, Speech and Signal
  Processing (ICASSP)}, 2022, pp. 6487--6491.

\bibitem{li22c_interspeech}
A.~Li, G.~Yu, C.~Zheng, and X.~Li, ``{TaylorBeamformer: Learning All-Neural
  Beamformer for Multi-Channel Speech Enhancement from Taylor’s Approximation
  Theory},'' in \emph{Proc. Interspeech 2022}, 2022, pp. 5413--5417.

\bibitem{9747528}
M.~M. Halimeh and W.~Kellermann, ``Complex-valued spatial autoencoders for
  multichannel speech enhancement,'' in \emph{ICASSP 2022 - 2022 IEEE
  International Conference on Acoustics, Speech and Signal Processing
  (ICASSP)}, 2022, pp. 261--265.

\bibitem{ren2021causal}
X.~Ren, X.~Zhang, L.~Chen, X.~Zheng, C.~Zhang, L.~Guo, and B.~Yu, ``A causal
  u-net based neural beamforming network for real-time multi-channel speech
  enhancement.'' in \emph{Interspeech}, 2021, pp. 1832--1836.

\bibitem{lee2021inter}
D.~Lee, S.~Kim, and J.-W. Choi, ``Inter-channel conv-tasnet for multichannel
  speech enhancement,'' \emph{arXiv preprint arXiv:2111.04312}, 2021.

\bibitem{9746054}
M.~Hao, J.~Yu, and L.~Zhang, ``Spatial-temporal graph convolution network for
  multichannel speech enhancement,'' in \emph{ICASSP 2022 - 2022 IEEE
  International Conference on Acoustics, Speech and Signal Processing
  (ICASSP)}, 2022, pp. 6512--6516.

\bibitem{barker2015third}
J.~Barker, R.~Marxer, E.~Vincent, and S.~Watanabe, ``The third {‘CHiME’}
  speech separation and recognition challenge: Dataset, task and baselines,''
  in \emph{2015 IEEE Workshop on Automatic Speech Recognition and Understanding
  (ASRU)}, 2015, pp. 504--511.

\bibitem{liu2020multichannel}
C.-L. Liu, S.-W. Fu, Y.-J. Li, J.-W. Huang, H.-M. Wang, and Y.~Tsao,
  ``Multichannel speech enhancement by raw waveform-mapping using fully
  convolutional networks,'' \emph{IEEE/ACM Transactions on Audio, Speech, and
  Language Processing}, vol.~28, pp. 1888--1900, 2020.

\bibitem{li2019multichannel}
X.~Li and R.~Horaud, ``Multichannel speech enhancement based on time-frequency
  masking using subband long short-term memory,'' in \emph{2019 IEEE Workshop
  on Applications of Signal Processing to Audio and Acoustics (WASPAA)}, 2019,
  pp. 298--302.

\bibitem{vincent2006performance}
E.~Vincent, R.~Gribonval, and C.~F{\'e}votte, ``Performance measurement in
  blind audio source separation,'' \emph{IEEE Transactions on Audio, Speech,
  and Language Processing}, vol.~14, no.~4, pp. 1462--1469, 2006.

\bibitem{rix2001perceptual}
A.~RIX, ``Perceptual evaluation of speech quality (pesq)-a new method for
  speech quality assessment of telephone networks and codecs,'' in \emph{Proc.
  IEEE International Conference on Acoustics, Speech, and Signal Processing,
  2001}, 2001, pp. 73--76.

\bibitem{taal2011algorithm}
C.~H. Taal, R.~C. Hendriks, R.~Heusdens, and J.~Jensen, ``An algorithm for
  intelligibility prediction of time--frequency weighted noisy speech,''
  \emph{IEEE Transactions on Audio, Speech, and Language Processing}, vol.~19,
  no.~7, pp. 2125--2136, 2011.

\end{thebibliography}


\begin{thebibliography}{00}
\bibitem{b1} G. Eason, B. Noble, and I. N. Sneddon, ``On certain integrals of Lipschitz-Hankel type involving products of Bessel functions,'' Phil. Trans. Roy. Soc. London, vol. A247, pp. 529--551, April 1955.
\bibitem{b2} J. Clerk Maxwell, A Treatise on Electricity and Magnetism, 3rd ed., vol. 2. Oxford: Clarendon, 1892, pp.68--73.
\bibitem{b3} I. S. Jacobs and C. P. Bean, ``Fine particles, thin films and exchange anisotropy,'' in Magnetism, vol. III, G. T. Rado and H. Suhl, Eds. New York: Academic, 1963, pp. 271--350.
\bibitem{b4} K. Elissa, ``Title of paper if known,'' unpublished.
\bibitem{b5} R. Nicole, ``Title of paper with only first word capitalized,'' J. Name Stand. Abbrev., in press.
\bibitem{b6} Y. Yorozu, M. Hirano, K. Oka, and Y. Tagawa, ``Electron spectroscopy studies on magneto-optical media and plastic substrate interface,'' IEEE Transl. J. Magn. Japan, vol. 2, pp. 740--741, August 1987 [Digests 9th Annual Conf. Magnetics Japan, p. 301, 1982].
\bibitem{b7} M. Young, The Technical Writer's Handbook. Mill Valley, CA: University Science, 1989.
\end{thebibliography}

\end{document}